# On broad iron K$\alpha$ lines in Seyfert 1 galaxies

A.C. Fabian[1], K. Nandra[1], C.S. Reynolds[1], W.N. Brandt[1], C. Otani[2], Y. Tanaka[3,4], H. Inoue[4], K. Iwasawa[1]

1. Institute of Astronomy, Madingley Road, Cambridge CB3 0HA
2. RIKEN, Institute of Physical and Chemical Research, Hirosawa, Wako, Saitama 351-01, Japan
3. Max-Planck-Institut fur Extraterrestrische Physik, D85470 Garching, Germany
4. Institute of Space and Astronautical Science, 3-1-1 Yoshinodai, Sagamihara, Kanagawa 229, Japan

**ABSTRACT**

The X-ray spectrum obtained by Tanaka et al from a long observation of the active galaxy MCG$-6-30-15$ shows a broad iron K$\alpha$ line skewed to low energies. The simplest interpretation of the shape of the line is that it is due to doppler and gravitational redshifts from the inner parts of a disk about a massive black hole. Similarly broad lines are evident in shorter observations of several other active galaxies. In this paper we investigate other line broadening and skewing mechanisms such as Comptonization in cold gas and doppler shifts from outflows. We have also fitted complex spectral models to the data of MCG$-6-30-15$ to see whether the broad skewed line can be mimicked well by other absorption or emission features. No satisfactory mechanism or spectral model is found, thus strengthening the relativistic disk line model.

## 1 INTRODUCTION

A long ASCA X-ray observation has resolved the iron K$\alpha$ emission line in the Seyfert 1 galaxy MCG$-6-30-15$ and shown that it is broad and skewed to the low energy end of the spectrum (Tanaka et al 1995). There is little line emission above the rest energy of 6.4 keV, close to where the line peaks, and considerable emission down to below 5 keV. The immediate interpretation of this feature is that the line originates by fluorescence in the very inner parts of an accretion disk, $3R_{\rm s} \leq R \leq 10R_{\rm s}$, about a massive black hole of Schwarschild radius $R_{\rm s}$ (Tanaka et al 1995). The line shape and skewness are due to the combined action of doppler shifts and gravitational redshift from matter moving in directions at most 30 deg from perpendicular to the line of sight. Here we investigate alternative interpretations in order to assess the strength of the inner accretion disk model.

Previous shorter ASCA observations had resolved the iron line in MCG$-6-30-15$ (Fabian et al 1994a) and other Seyfert 1 galaxies; NGC5548 (Fabian et al 1994b; Mushotzky et al 1995), IC4329A (Mushotzky et al 1995), NGC3227 (Ptak et al 1994), Fairall 9 (Otani et al 1995), Mrk1040 (Reynolds, Fabian & Inoue 1995) and 3C390.3 (Eracleous et al 1995). A resolved line is also seen in the Seyfert 1.9/2 galaxy IRAS18325 $-$ 5926 (Iwasawa et al 1995). These results showed that the line is broad, with FWHM ranging from 10,000 to 50,000 $\rm km\,s^{-1}$ , but were not adequate to define the skewness of the line. The line in NGC 7469 is not broad (Guainizzi et al 1994), unless some weak emission at 6.7 – 6.9 keV is part of the line. George et al (1995) find no evidence for broadening of the line in NGC3783.

The discovery of X-ray reflection in Seyfert 1 galaxies with GINGA spectra (Pounds et al 1990; Matsuoka et al 1990; see also Nandra & Pounds 1994) has already provided strong evidence for the presence of cold matter subtending about $2\pi\,$sr at the X-ray source (George & Fabian 1991; Matt, Perola & Piro 1991; Matt et al 1992). An accretion disk would realize such a geometry. A clear test which appears to be satisfied by the ASCA observations is that the fluorescent iron line is broad and skew due to doppler and gravitational shifts from the deep gravitational potential.

The iron line in the long ASCA spectra of MCG$-6-30-15$ is fit well (Tanaka et al 1995) by the line expected from the inner parts of a disk with an inclination of 30 deg orbiting about either a Schwarzschild (Fabian et al 1989) or Kerr (Laor 1991) black hole.

## 2 ALTERNATIVE MECHANISMS FOR THE PRODUCTION OF A BROAD, SKEW IRON LINE

In this section we investigate alternatives ways to generate the broad, skewed iron line seen in MCG$-6-30-15$. The obvious possibilities are Comptonization, jets and/or outflows, photoelectric absorption and resonance absorption. As we discuss, none provides a satisfactory explanation for the observed feature.

### 2.1 Comptonization

Czerny, Zbyszewska & Raine (1991) suggested Comptonization as the broadening mechanism for the possible broad line seen in Ginga spectra of NGC3227 (George et al 1991). The resolved ASCA data however allow us to place very strong constraints on any Comptonization model. As discussed by Tanaka et al (1995) and Mushotzky et al (1995) the fact that the line is skewed to low energies means that any Comptonizing gas must be essentially cold. The redshift of the line is then due to successive downscattering requiring a Thomson depth for the scattering medium, $\tau$, of at least 3 so that $\sim 10$ scatterings can shift the line centroid by at least 800eV. Indeed since the blue wing of the line is less than 0.2 keV wide, there can be little thermal dispersion to the line, meaning that both the temperature of the gas $kT < 0.25\,\mathrm{keV}$ and the whole line spread of at least 1.5 keV must be due to electron recoil accumulated over many scatterings. $\tau$ then needs to be close to 5 (see Pozdnyakov, Sobol & Sunyaev 1979 for simulations of such Comptonization).

Since, in a Comptonization model, most of the X-rays must pass through gas in which $kT < 0.25\,\mathrm{keV}$ and $\tau \sim 5$, there will be a marked effect on the spectral continuum. A break should be seen at $511/\tau^2\,\mathrm{keV} \approx 20\,\mathrm{keV}$, contrary to observations (Zdziarski et al 1995). Also there will be large photoelectric edges apparent from such gas unless it is extremely highly ionized. Using the argument in Mushotzky et al (1995) and the photoionization calculations of Kallman & McCray (1982), we can combine the required ionization parameter limit $\xi = L/nR^2 > 10^4$ (in order that iron is completely stripped) with the required column density ($\sim 7.5 \times 10^{24}\,\mathrm{cm}^{-2}$) to see that in MCG$-6-30-15$ ($L \sim 10^{43}\,\mathrm{erg\,s}^{-1}$) then the radius of the gas $R < 10^{14}\,\mathrm{cm}$ (the radius given in Mushotzky et al 1995 is incorrect). This means that the iron line originates from a region of less than $50R_s$ in which case there should be large doppler motions overwhelming the effects of Comptonization. The argument is strengthened further after reworking with the recombination rate for iron (Shull & Van Steenburg 1982) appropriate for $kT < 0.25\,\mathrm{keV}$ (the photoionization equilibrium temperature in the work of Kallman & McCray 1982 is considerably higher).

It is moreover difficult to understand why the scattering gas should be below the probable Compton temperature in the source.

We note that any Comptonization off distant matter may involve bound electrons, presumably predominantly of hydrogen. The probability that the scattering involves any significant nucleus recoil is very low (much less than 1 per cent; Gorshkov, Michailov & Sherman 1973) so the expressions for simple free-electron assumed above are appropriate.

### 2.2 Jets and outflows

Many Active Galactic Nuclei (AGN) show collimated jets although none are clearly seen in the Seyfert 1 galaxies mentioned here. MCG$-6-30-15$ is radio quiet and unresolved by the VLA (Ulvestad & Wilson 1984). In order to produce a redshifted broad line a spread of velocities are required which are mostly perpendicular to the line of sight. Perhaps a jet accelerating perpendicular to the line of sight or within $R \sim 10R_s$ might account for the observed features, with the redshift dominated by transverse doppler and gravitational redshifts, respectively. There is a further problem in producing the line within the jet; the jet would have to be fanshaped (again mostly perpendicular to the line of sight) in order that the line emission is excited by X-ray illumination. The lack of any strongly blue shifted lines in the heterogenous sample of objects observed so far with ASCA (and the larger Ginga sample) further argues against any jet model.

A more general accelerating outflow of Thomson-thick cold clouds, irradiated by a central source, provides a better alternative model. We might then only detect the iron line from receding clouds with velocities $\sim 100,000\,\mathrm{km\,s}^{-1}$ since the emission-reflecting face of the approaching clouds is obscured from us by the bulk of such clouds. Such a model must have an *ad hoc* velocity distribution in order to account from the observed line and also must obscure at least half the sky subtended at the source in order to account for the line strength. All clouds must be optically thick in order that any blue wing is as small as observed.

A major problem with the outflow model, apart from its ad hoc nature, is that MCG$-6-30-15$ has a warm absorber. The OVII edge in that indicates that an flow along our line of sight is less than about 5,000 $\mathrm{km\,s}^{-1}$ (Fabian et al 1994a), which seriously conflicts with an outflow of about 100,000 $\mathrm{km\,s}^{-1}$.

We note here that a 22.5 ks ROSAT HRI observation of MCG$-6-30-15$, after correction using the algorithm of Morse (1994), does not show any strong extended soft X-ray emission (Fig. 1), such as expected from a powerful outflow.

### 2.3 Photoelectric and resonance line absorption, and other spectral features

A very broad emission line could appear skewed if the blue wing were absorbed. Photoelectric absorption is most unlikely to explain the feature observed in MCG$-6-30-15$ since the blue side of the line drops very sharply at about 6.6 keV. This is well below the lowest photoelectric edge of iron at 7.1 keV so significant doppler or other shifts are then required. In order that the low energy part of the spectrum is not completely absorbed (at least by transitions of iron L), the iron must be ionized, thus increasing the energy of the edge and requiring larger velocity (or other) shifts, comparable to those required in the simple accretion–disk–reflection model. It is of course most plausible that the sharp drop is close to the rest energy of the feature which produces it, rather than at an extreme point in a varying shift. This further favours an emission line interpretaion over an absorption edge one.

Compton reflection from the obscuring torus has been suggested as the origin of (at least part) of the emission line and reflection hump in Seyfert galaxies (Ghisellini, Haardt & Matt 1994; Krolik et al. 1994). However, these models predict a narrow emission line. Line–of–sight material could also produce small emission lines between 6.4–6.9 keV, depending on the ionization state and could be blended into an apparently “broad” feature. Furthermore, our knowledge of the shape of the continuum is not perfect, and the power–law we have assume could conceivably be modified by other processes. We now investigate whether a complex continuum or line blends could mimic the broad line–profile observed in MCG−6 − 30 − 15. Here we consider several additional components which could plausibly be present in the spectrum.

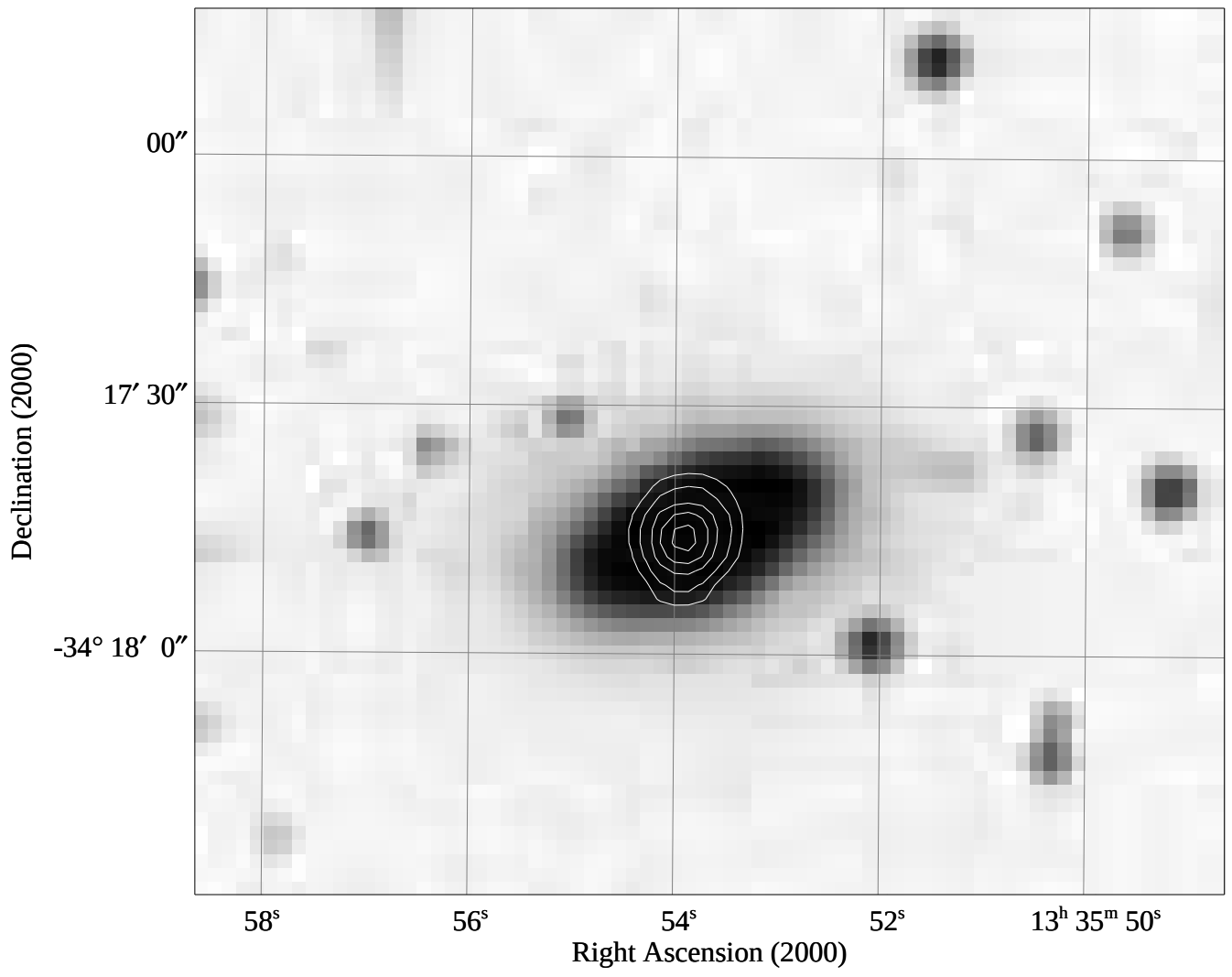

**Figure 1.** Contours of the Morse-corrected ROSAT HRI image overlaid on the UK Schmidt southern sky survey J plate. Contours are at 11.3, 22.5, 45.1, 67.7 and 90.3 per cent of the maximum pixel value. Note the good positional coincidence, compact nature of the X-ray source, and lack of secondary X-ray sources.

Our starting point for an investigation of these possibilities is a power–law with a Gaussian line with rest energy 6.4 keV and narrow compared to the resolution of ASCA. It is our intention thereby to test whether a narrow emission line (or a blend of narrow lines), when combined with complex continuum components, can account for the observed spectrum, mimicking the broad–profile. We note first that this power–law plus narrow line model is a considerably worse fit to the data ($\chi^2$=748/680 d.o.f.) than the model with the disk line ($\chi^2$=655.7/676 d.o.f). Except where stated, the fits are carried out to the 3–10 keV data from the ASCA data of MCG−6 − 30 − 15 obtained with both SIS0 and SIS1 detectors.

(i) Enhanced reflection continuum: Adding a reflection continuum and allowing the solid angle of the reflector to be a free parameter does improve the fit, but only to $\chi^2$ = 732.6/679 d.o.f. Furthermore the best–fit value for the solid angle is $\Omega/2\pi = 5.0$. It is difficult to conceive of any geometry which could produce such a large reflection continuum with such a small line (EW $\sim$ 35 eV).

(ii) Iron absorption edge: Adding an absorption edge, restricted in energy between the (physically appropriate) values for iron of 7.1-9.29 keV (assuming no relativistic shifts) improves the fit to only $\chi^2$=734/678 d.o.f. The best–fit edge has an energy of 7.57 keV and optical depth of $\tau = 0.16$. If the energy of the edge is unrestricted then a better fit is obtained at 6.8 keV with a depth of 0.2 ($\chi^2$=708/678). As already mentioned, an interpretation based on this result requires that the edge is redshifted and so requires relativistic effects, particularly if an edge due to highly ionized iron is assumed ( so that there is no L-shell absorption at lower energies).

(iii) Line blends: Adding another, narrow line to the model with free energy, but restricted to the physical range of 6.4-6.9 keV (again, assuming no relativistic shifts operate) improves the fit to $\chi^2$=735.7. Adding a further Gaussian improves the fit negligibly ($\Delta\chi^2$=0.6).

(iv) Partial covering: An absorber which partially covers the source improves the fit to $\chi^2$=689/678 d.o.f. There are however serious problems with this interpretation since the required photon index is steep (photon index $\Gamma = 2.5$) and most of the significant residuals occur between 6 and 7 keV. Extending the energy range of the fit down to 0.5 keV shows that a flatter index, $\Gamma \sim 1.95$, is required and leaves residuals similar in shape to a diskline.

(v) An exponential cutoff in the power–law: This improves the fit to $\chi^2 = 745.6$ with cutoff energy $E_{\rm c}$ = 21.9 keV, inconsistent with other data (Zdziarski et al 1995).

(vi) Additional continuum components: Adding a black body, whose broad shape could mimic the observed "bump" at $\sim$ 5.5 keV, gives $\chi^2$=700/688 d.o.f. Note that variability of the power-law index during the observation will make the overall spectrum concave, opposite in shape to that of the disk line. The pointlike nature of the ROSAT HRI source and absence of any other point source which could confuse the ASCA data (Fig. 1) strongly reduce the possibility of any non-nucleus continuum contamination to the spectrum of MCG−6 − 30 − 15.

As a final test of a "continuum conspiracy" we have tried a model with combinations of *all* of the above components. The best fit with a power–law, blackbody, free reflection component, iron absorption edge, partial covering and two gaussian lines has $\chi^2$ = 678.1/670 d.o.f, *still* worse that of the disk line and with 6 more free parameters. We note that the reflection component and black body do not improve this fit at all and that the edge does so negligibly ($\Delta\chi^2$=2).

We do know from oxygen edges in the spectrum of MCG−6−30−15 that there is a warm absorber along the line of sight with a column density of about $10^{22}\,{\rm cm}^{-2}$ (Fabian et al 1994a; Reynolds et al 1995). The photoelectric edge from that absorbing gas is undetectable. Resonance line absorption in this gas should also be negligible, absorbing an equivalent width of at most 20 eV (Matt 1994).

A more highly ionized gas with a larger column ($\gtrsim 10^{23}\,{\rm cm}^{-2}$) in which oxygen is mostly fully stripped and iron helium-like or hydrogenic could produce greater resonance absorption if there is a velocity spread exceeding about 10,000 ${\rm km\,s^{-1}}$. This is problematic since the observed oxygen warm absorber has a velocity $< 3000\,\mathrm{km\,s^{-1}}$ (Fabian et al 1994a). Furthermore the gas parameters relating to both resonance and photoelectric absorption would have to be in a narrow range in order that continuum absorption is not apparent in the observed spectrum. The emission line itself would also have to be of much larger equivalent width than the observed value ($\sim 380\,\mathrm{eV}$).

## 3 DISCUSSION

We have not found any valid alternative to the accretion disk model for the observed, skewed emission line. Variability of the line and reverberation mapping would be conclusive evidence for the model but may require a much larger count rate than is possible with ASCA.

The results from a wide sample of AGN should also help to define the model since a range of inclination angles should be found. Curiously most have an inclination around 30 deg (but note that IRAS18325 − 5926 has an inclination of 40 – 50 deg; Iwasawa et al 1995) . This could in part be a selection effect due to a) the equivalent width of the line decreasing and b) the line width increasing as the inclination increases. The line would then become very difficult to determine, especially since the underlying continuum, which carries most of the flux, would be difficult to define. Also at high inclinations (perhaps exceeding 45 deg or so) the obscuring torus required by the unified Seyfert model (see Antonucci 1993 and references therein) would obscure the direct X-ray source, at best making the continuum even more difficult to define. We note too that above 60 deg the accretion disk corona (assumed to have a vertical Thomson depth $sim 0.3$ and a temperature $kT \sim 150\,\mathrm{keV}$ similar to that required for Cygnus X-1; Haardt et al 1993) would severely Comptonize, and thus destroy, sharp features in the reflection spectrum.

A further reason why 30 deg is emerging as a common angle may be that the line is usually observed to peak around the rest energy (6.4 keV) and does not extend far above that energy. A line of rest energy $E$ from radius $R$ in a relativistic disk approximately covers the energy band $E(1 - 3R_{\mathrm{s}}/4R \pm \sqrt{R_{\mathrm{s}}/2R}\sin i)$, where $i$ is the inclination. The maximum upper cutoff energy to the line profile from a disk is then given by $E_{\max} = E(1 + \sin i^2/6)$ (this is a fair approximation to the correct profile from a relativistic disk, Fabian et al 1989, provided that $i < 50$ deg). For the cold iron line $E = 6.4\,\mathrm{keV}$ and $E_{\max} = 6.7\,\mathrm{keV}$ when $i = 32$ deg (similar to MCG$-6-30-15$) and $E_{\max} < 7\,\mathrm{keV}$ for $i < 50$ deg. The long spectra of MCG$-6-30-15$ are inconsistent with a significant narrow 6.4 keV component, but the data on other AGN are not yet of the quality to rule out any strong contribution. It is possible that MCG$-6-30-15$ does have an inclination close to 30 deg but that several of the less well-observed objects have higher inclinations which are spectrally masked by a strong narrow component at 6.4 keV originating from larger radii in the source (e.g. reflection from a torus).

The disk model for MCG$-6-30-15$ requires that the line be strongly concentrated to the inner parts of the disk. This means that the primary hard X-ray source which excites the fluorescence must be close to the disk so that it cannot irradiate much of the outer radii. The uncertainty on the outer disk radius obtained for MCG$-6-30-15$ ($2.5R_{\mathrm{s}}$; Tanaka et al 1995) should roughly be the limit on the height of hard source above the disk. The results also imply that the velocity vector of most of the disk material is inclined to the line of sight by more than 60 deg, so constraining the opening angle of any thick, or blobby, disk (or collection of orbiting clouds). We note that the Bardeen-Petterson (1975) effect around a spinning black hole will tend to force matter into the equatorial plane of the hole, thus giving it a disk-like configuration.

If the hard source shares the rotation of the disk, then it will be significantly beamed. If the emission is produced by flares above the disk, then a flare on the approaching (blue) side may appear two or more times brighter than a similar flare on the receding (red) side (the effect of $(1+z)^4$, where $z$ is the redshift; Luminet 1979). This means that the response of the line to continuum changes may be complex. The blue wing may be much better correlated with continuum changes than the red wing, since the continuum is dominated by flares on the blue wing.

We note that the Galactic Black Hole Candidates (BHC) tend to show broad, smeared edges beginning about 7.1 keV, rather than broad emission lines (e.g. Ebisawa 1991, Tanaka 1992). This can be explained by the disk in BHC being fairly highly ionized (Ross, Fabian & Brandt 1995); the much smaller mass of the black hole in these systems causes the disk to be hotter. They may also operate closer to the Eddington limit. The reflection continuum is then significant at all energies apart from in the iron absorption edge, which by contrast becomes detectable. Much of the line in BHC may be rendered undetectable by Auger destruction (Ross et al 1995).

The equivalent width of the relativistic disk line obtained with the ASCA data is higher than the mean value for the GINGA observations of 140 eV (Nandra & Pounds 1994). The line width was however restricted to be narrow. Refitting the GINGA data with the disk line model with parameters fixed at those found by Tanaka et al (1995) yields an equivalent width $\sim 250\,\mathrm{eV}$, compatible with the ASCA observations.

In summary, we have explored a number of alternative interpretations for the broad, skew iron line seen in MCG$-6-30-15$ and other active galaxies and find that none is satisfactory. This strengthens the plausibility that the line is caused by fluorescence from a relativistic disk.

## 4 ACKNOWLEDGEMENTS

ACF thanks the Royal Society and British Council, WNB thanks the United States National Science Foundation and the British Overseas Research Studentship programme, and KI thanks the JSPS and the British Council for support.